\documentclass[journal]{IEEEtran}
\ifCLASSINFOpdf
\else
\fi

\usepackage{graphicx}
\usepackage{float}
\usepackage{longtable}
\usepackage{tabularx}
\usepackage{color}
\usepackage{caption}
\usepackage{subcaption}
\usepackage{gensymb}
\usepackage{multirow}
\usepackage{stfloats}
\usepackage{float}

\begin{document}
\title{ChatSUMO: Large Language Model for Automating Traffic Scenario Generation in Simulation of Urban MObility}

\author{Shuyang Li,
        Talha Azfar,
        and~Ruimin Ke,~\IEEEmembership{Member,~IEEE}% <-this % stops a space
\thanks{S. Li was with the Department
of Civil and Environmental Engineering, University of Michigan, Ann Arbor,
MI, 48109.}% <-this % stops a space
\thanks{Talha Azfar and Ruimin Ke (e-mail: ker@rpi.edu) are with the Department of Civil and Environmental Engineering, Rensselaer Polytechnic Institute, Troy, NY, 12180.}% <-this % stops a space
\thanks{Manuscript received xxx, 2024; revised xxx.}}


\markboth{IEEE Journal}%
{Shell \MakeLowercase{\textit{et al.}}: Bare Demo of IEEEtran.cls for IEEE Journals}



\maketitle

\begin{abstract}
Large Language Models (LLMs), capable of handling multi-modal input and outputs such as text, voice, images, and video, are transforming the way we process information. Beyond just generating textual responses to prompts, they can integrate with different software platforms to offer comprehensive solutions across diverse applications. In this paper, we present ChatSUMO, a LLM-based agent that integrates language processing skills to generate abstract and real-world simulation scenarios in the widely-used traffic simulator - Simulation of Urban MObility (SUMO). Our methodology begins by leveraging the LLM for user input which converts to relevant keywords needed to run python scripts. These scripts are designed to convert specified regions into coordinates, fetch data from OpenStreetMap, transform it into a road network, and subsequently run SUMO simulations with the designated traffic conditions. The outputs of the simulations are then interpreted by the LLM resulting in informative comparisons and summaries. Users can continue the interaction and generate a variety of customized scenarios without prior traffic simulation expertise. For simulation generation, we created a real-world simulation for the city of Albany with an accuracy of 96\%. ChatSUMO also realizes the customizing of edge edit, traffic light optimization, and vehicle edit by users effectively.
\end{abstract}


% \fi
\IEEEpeerreviewmaketitle

\section{Introduction}
\IEEEPARstart{T}{he} increasing complexity of modern transportation systems, with diverse vehicle types and traffic patterns, poses significant challenges for traffic management and forecasting~\cite{cui2019traffic, ke2020smart}. This complexity not only escalates transportation costs but also contributes to environmental pollution. The need for improved traffic planning and operation has led to a surge in studies focused on optimizing transportation systems, e.g., the strategic reconstruction of road infrastructure~\cite{wang2023machine, Dorokhin_2020}. Traffic simulation has emerged as a powerful tool for modeling current traffic scenarios, predicting future conditions, and mitigating negative impacts, all while reducing the costs associated with real-world traffic planning implementations~\cite{6847430}. Among these tools, SUMO (Simulation of Urban MObility) stands out as a versatile, open-source platform for traffic simulation, used widely for urban mobility research, operations, and planning~\cite{krajzewicz2002sumo}.

Despite its effectiveness, creating traffic simulation scenarios is a time-consuming process that requires specialized traffic-related knowledge~\cite{azfar2022efficient}.
Most mainstream simulation software demands that users define networks, vehicles, routes, and other parameters, which poses a significant barrier to entry for beginners who lack professional expertise or even for experts but without the experience in the certain software~\cite{ejercito2017traffic,nguyen2021overview}. These users often seek quick access to modeling results without the need for extensive setup and configuration.

The advent of Large Language Models (LLMs), trained on vast datasets, offers a promising solution by facilitating a more intuitive human-machine interaction. LLMs can interpret a wide range of inputs, including text, images, and videos, and generate corresponding outputs~\cite{chang2024survey}. SUMO, a popular open-source traffic simulation software, requires users to either code networks from scratch or convert them from other platforms~\cite{8569938}. Additionally, users must manually define traffic flows or run Python scripts with specific parameters, adding to the software's learning curve~\cite{wu2024nextgptanytoanymultimodalllm}.

To address these challenges, we present ChatSUMO, a cutting-edge LLM-based assistant designed to streamline the use of SUMO simulations. Powered by the Llama 3.1 model~\cite{touvron2023llama,llama_2024}, ChatSUMO enables users to generate and modify traffic simulation scenarios through simple textual inputs. This framework transforms user descriptions into executable SUMO simulations using Python scripts, effectively lowering the barrier for those without specialized knowledge. ChatSUMO operates by leveraging a multi-module architecture to facilitate user interaction and simulation generation. The system begins with an Input Module, which processes user inputs and converts them into relevant keywords. These keywords are then used by the Simulation Generation Module to create either abstract or real-world traffic scenarios in SUMO. Users can customize these scenarios using the Customization Module, which supports a range of modifications, including edge and lane edits, traffic light optimization, and vehicle route adjustments. The Analysis Module interprets the simulation outputs, providing detailed reports on traffic density, travel time, emissions, and more. The contributions are summarized as follows:
\begin{itemize} 
    \item We propose a novel LLM-based agent capable of transforming textual descriptions into SUMO simulation scenarios. This allows users to bypass the need for extensive traffic simulation knowledge.
    \item ChatSUMO streamlines the process of generating and modifying simulations, making it accessible to users of all expertise levels.
    \item By leveraging advanced language processing capabilities, ChatSUMO provides an intuitive interface for traffic simulation, offering real-time insights and dynamic adjustments.
\end{itemize}

The rest of the paper is structured as follows: we first review related literature on traffic simulation and the application of LLMs in this domain. We then detail the methodology behind ChatSUMO's design and functionality, followed by an experimental evaluation of its performance in generating and modifying traffic simulations. Finally, we discuss potential applications and conclude with future work directions aimed at enhancing the system's capabilities.

\section{Literature Review}
%transportation engineers, planners, administrators
LLM research is focused on enhancing natural language processing objectives including text classification, language inference, and semantic understanding. While they face challenges in reasoning, ethics, and conflict resolution, they have proven to be excellent tools for summarization, contextual comprehension, and question answering \cite{chang2024survey}. Modern LLMs have undergone training on vast quantities of data and their behavior has been fine tuned by human feedback, such that the most competitive models are very good at following instructions and remaining focused towards specified tasks \cite{ouyang2022training}. 
LLMs show promise in the enhancement of education where precise answers or subject matter experts may not be easily accessible \cite{KASNECI2023102274}. 

In the transportation field some LLM related work has emerged recently, focusing primarily on safety. TrafficSafetyGPT \cite{zheng2023trafficsafetygpt} finetuned Llama on a custom dataset curated from NSTHA Model Minimum Uniform Crash Criteria guidelines, FHWA Highway Safety Manual, and ChatGPT generated data. The model learned domain specific concepts allowing it to accurately answer challenging transportation safety questions with concise answers. ChatScene \cite{zhang2024chatscene} was developed to generate safety-critical scenarios for autonomous vehicles as text descriptions which are then broken down into sub-descriptions that can be used to instantiate the scenario in CARLA. A database of scene components and descriptions was created that enabled ChatScene to assemble scene scripts from LLM output. AccidentGPT \cite{accidentgpt2024} combines scene perception and trajectory prediction using computer vision on camera views from multiple vehicles and roadside units for environmental understanding and collision avoidance. GPT4 based reasoning module is then used to provide proactive cues for human drivers and traffic management authorities. It also stores key moments and uses it for later analysis to improve future autonomous driving decisions. Traffic Performance GPT (TP-GPT) proposes an intelligent chatbot designed to aid in transportation analytics. The TP-GPT utilizes LLMs to generate accurate SQL queries and interpret traffic data, leveraging a real-time database of traffic information~\cite{wang2024traffic}.

Language models have been used in combination with computer vision for scene understanding for autonomous driving in a variety of techniques \cite{zhou2024}. ADAPT (Action-aware Driving cAPtion Transformer) \cite{jin2023} provides an innovative end-to-end transformer-based approach for generating action narration and reasoning in self-driving vehicles. ADAPT employs multi-task joint training to bridge the gap between driving action captioning and control signal prediction. ChatGPT was used as a co-pilot for assisted driving in \cite{wang2023} by converting vehicle telemetry, road state, human intention, and descriptions of the available controllers into a combined prompt. The response from the LLM determines the course of action most appropriate for those conditions. The system can switch between aggressive and gentle controllers, and handle lane changes and overtaking. The DiLu framework \cite{wen2023dilu} incorporates GPT based reasoning and reflection modules to perform decision making for an autonomous vehicle and has the ability to learn continuously. The system is able to use LLM common sense chain of thought reasoning from prompts tailored to the scenario which generates the final decision. Meanwhile the decision sequences stored to memory can be reflected upon by the LLM to find mistakes and correct them. Similarly, LanguageMPC \cite{sha2023languagempc} used an LLM for high level autonomous driving decision making, converting text descriptions to mathematical representations to be used by the model predictive controller. It was able to handle multi-vehicle coordinated control by generating a convoy level decision that each vehicle interprets according to its internal state. BEVGPT is a generative pre-trained model that integrates driving scenario prediction, decision-making, and motion planning into a minimalist autonomous driving framework using only bird's-eye-view images, outperforming previous methods in key metrics and pioneering long-term BEV image generation for autonomous driving~\cite{wang2023bevgpt}.

Microscopic traffic simulations such as VISSIM, SUMO, and MATSim are the basis of planning and optimization studies for traffic networks \cite{diallo2021comparative} and a few recent works have incorporated LLMs with microsimulation tools. PromptGAT \cite{da2024prompt} leverages LLM inference to understand how weather conditions, traffic states, and road types influence traffic dynamics, which is used to inform policy in reinforcement learning for traffic signal control. This additional information about real-world conditions helps to reduce the simulation to reality gap. In a similar vein, language assisted traffic light control in \cite{wang2023} employ LLM to understand the traffic observations and recommended action from reinforcement learning, which then generates a justification for the action using chain of thought reasoning. Anomalous traffic conditions like blockages, and the presence of emergency vehicles are some of the factors the LLM takes into consideration before selecting the appropriate action. In \cite{pmlr-v229-zhong23a}, natural language queries are translated into differentiable loss functions for specified vehicle trajectories in order to facilitate scenario based traffic simulations. These scenarios include car following and collision trajectories for a few vehicles and compare them to ground truth from nuScenes dataset. There have also been advances in using LLM for microscopic traffic behavior modeling, such as in~\cite{chen2024genfollower}, Chen et al. proposes a LLM-based method for car following behavior modeling; however, they do not necessarily include microscopic traffic simulations.

\section{Methodology}
\subsection{Overview}
The overview of ChatSUMO, as Figure 1, presents a structure of the proposed system. The framework is designed to assist traffic simulation generation. ChatSUMO integrates advanced chat model capabilities into the SUMO platform to enhance the efficiency and accuracy of traffic simulation and management. This integration leverages the power of the GPT model to simulate, modify, and analyze traffic scenarios, providing real-time insights and dynamic adjustments. Our methodology begins by leveraging the LLM for user input which converts to relevant keywords needed to run python scripts. These scripts are designed to create an abstract network or convert specified regions into coordinates, fetch data from OpenStreetMap, transform it into a road network, and subsequently run in SUMO simulation with the designated traffic conditions. We use Llama 3 8B, an open-source model, to parse the inputs and provide a summary of the output. The user can then request another simulation with some modifications which create a different traffic condition. The LLM retains context for continued interaction. The core component of this methodology is the gpt-reasoning module, which is responsible for three critical modules: Input Module, Simulation Generation Module, Simulation Customization Module, and Simulation Analysis Module. 

\begin{figure*}
    \centering
    \includegraphics[width=1\textwidth]{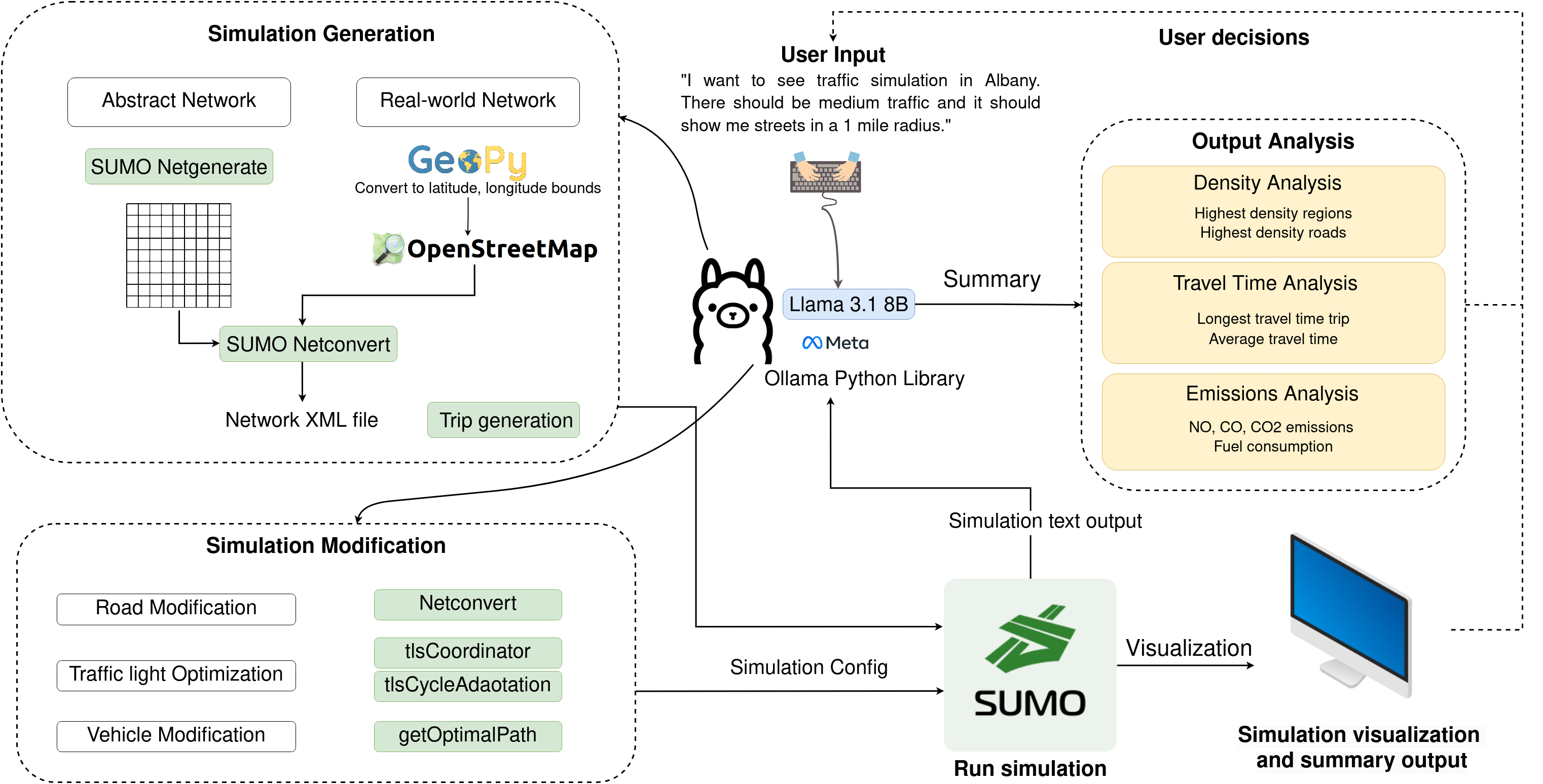}
    \caption{ChatSUMO Framework}
    \label{fig:framework}
\end{figure*}

\subsection{GPT-reasoning}
The GPT-reasoning framework serves as the core element of ChatSUMO's functionality. The process of our reasoning module is illustrated in Figure~\ref{fig:framework}, encompassing input, simulation generation, modification, analysis modules. We will now elaborate on the design of these modules, emphasizing their positive impact on increasing the human-machine interaction efficiency. In this work, we consistently use Llama 3.1 to decode the user input. In the reasoning module, it would first analyze the user's input which contains requirements (type of network, city for simulation, traffic volume) to generate the user's ideal simulation scenario. After running the initial simulation, the LLM would analysis the output, producing a report for the simulation. Then ChatSUMO asks the user what modification they want to utilize for optimization, and the modification module will comprehend the user's needs and modify the simulation scenario based on the specific commands. Finally, the LLM in analysis module analyzes the results of each simulation and the user can choose to compare the output from each step of modification, which includes information like traffic density, average travel time, emissions and fuel consumption.

\textbf{Input Module. }Input Module is header which deals with all the input information from users. In order to reduce the difficulty of creating traffic simulation scenarios, we have simplified the user input as much as possible, so that the user can create the desired simulation scenarios without entering traffic-professional descriptions using natural language. Based on the Meta llama3.1 model, we create SUMOInput as the traffic scenario identification model for analyzing users input. In this model, we customize it with specified some prompts as ``You are taking input and generate keywords for a transportation simulation. Analyze the user input and give a python dictionary with these keywords ...". To generate the initial simulation scenario, an example of user input can be: ``Generate a simulation in city Albany with a radius of 3miles, and the volume of traffic should be medium." After parsed by ChatSUMO, the natural language input would be transformed into a python dictionary. These python dictionaries usually conclude three parts: the input of decision, the input of types. the input of specific requirement. Regarding the decision input, it requires users to make a yes or no decision for the question. For the input of types, SUMOInput is expected obtain the type of the decision question (type of network, kind of modification). Specific requirement inputs are inputs that contains users detailed information about the simulation or modification (number of grids, which street to be removed). As the example above, the dictionary transformed from user input would be ``\{city: Albany, radius: 3 miles, traffic condition: medium\}". 

\textbf{Generation Module. }To generate the initial scenario of the Simulation, we build the simulation scenarios in SUMO by inputting the desired simulation scenarios from the user. Currently we can generate two types of simulation networks, including abstract scenarios (e.g., spider, grid networks) and real-world networks. For real-world networks, users can enter the name of the city, size of radius and the condition of traffic (e.g., light, medium, heavy).  After the user inputs a simulation scenario, the input would be analyzed and understood by Llama, and then it would be extracted to keywords as python dictionary. These keywords won't be directly transmitted to python, as ChatSUMO will analyze the user input and provides feedback on whether the input is sufficient to construct the simulation scenario. 

\begin{figure*}
    \centering
    \includegraphics[width=1\linewidth]{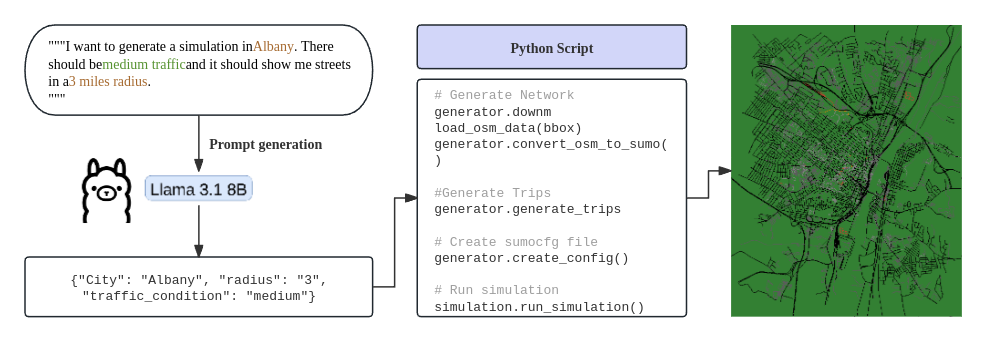}
    \caption{Simulation Generation}
    \label{fig:simulation generate}
\end{figure*}

After the simulation generate module gets sufficient information, these keywords would be processed by python script which executes commands for generating simulation. To download the OSM (OpenStreetMap) of the required region of city, ChatSUMO would execute ``osmGet.py" to obtain the osm map in predefined region. Then it would execute ``netconvert" commands which would convert the OSM map to network file in SUMO. After converting, it would utilize ``randomTrips.py" and generates random trips with converted network and required traffic volume. Finally, it would create the configuration file which can be executed by SUMO.

\textbf{Customization Module. } Based on the generated simulation, ChatSUMO supports the customization of simulation from the user's text descriptions by utilizing multiple customizing modules. After users enter their modification, the input module would analyze and match the keywords with some predefined customizing API. Through these apis, users can remove edges, optimize traffic light, and add vehicles to the simulation. Here are the details of implementing these APIs.

\textit{Edge and Lane Edit:} Users can make modifications to the roads in simulation by simply telling ChatSUMO which lanes to remove, e.g., "I want to remove Madison Avenue" or "I'd like to remove the first lane in Madison Avenue", thus validating some of the user's conjectures about traffic. To realize this function, ChatSUMO would first check whether the modified road is in the generated simulation, if so, the edit module obtains the modification type for the road. Then the module would extract the name of road as "Madison Avenue", and generate the terminal command for SUMO tool \textit{netconvert} through python script to modify the network. As user only input the general name of the removed street, multiple edges might be found, then ChatSUMO would ask for the user's decision which edge to be removed.

\textit{Traffic Light Offset}: Traffic light offsets are useful dealing with multiple traffic lights to increase the crossing efficiency of traffic flow. Users can enter commands like "I want to set offsets to all the traffic light in the simulation" to set all the traffic light in the simulation with offsets. With the traffic light offsets, intersections are capable of green wave control. To implement this function, once ChatSUMO receives the key word ``traffic light offset", it will generate terminal command to call \textit{tlsCoordinator.py} python script to modify the traffic-light offsets to coordinate them for the current traffic demand, and generate a tlsOffstes.add.xml which can be loaded into SUMO. 

\textit{Traffic Light Adaptation}: Users can enter command like "I want to set offsets to all the traffic light in the simulation" to optimize the traffic-light cycle in the simulation with  traffic light adaptation api. To implement this function, once ChatSUMO receives the command, it will call \textit{tlsCycleAdaptation.py} python script generate an additional newTLS.add.xml file to sumo configuration, which modifies the signal cycle length and the duration of green phases according to Websters formula to best accommodate a given traffic demand. 

\textit{Vehicle Generate}: The vehicle generate API is used to generate a vehicle with a given depart and arrival edge-pair. After the user entered the origin and destination, ChatSUMO would first whether these roads are contained in the network, or it would tell the user "Entered Roads are not in the current network". To generate route for the vehicle, the module would call \textit{getOptimalPath} to finds the optimal (shortest or fastest) path from depart edge to arrival edge by using Dijkstra's algorithm. Then, a vehicle with the assigned route would be add into the cityname.rou.xml file, which would be loaded into SUMO simulation later.

\textit{Vehicle Type Edit}: In the initial traffic simulation settings, both gas vehicles and electric vehicles are generated, and the proportion of them is 0.5 and 0.5. To change the proportion of vehicle types, users can utilize the vehicle type edit module. To implement the customization, ChatSUMO creates a vehicle type dictionary which stores the detail information for each vehicle type. After the user entered the modified proportion, ChatSUMO would utilize \textit{RandomTrips.py} to generate the new route file, including the customized vehicle proportion. 

\textbf{Analysis Module. } For analysis module, it will process data from the output xml file generated by simulation, and interpret them into analysis report, which involves density analysis, travel time analysis, emission analysis. Based on the output of simulation, ChatSUMO would identify the top 10 congested roads, average travel time, the emission of pollutants, including CO$_2$, CO, PMx, and the fuel consumption. Every time customization ends, the output of simulation would be stored into a database. Each time the user makes customization to the simulation, ChatSUMO would run another round of simulation. When the simulation finished, ChatSUMO asks the user if they want to make a comparison with any previous simulation, giving a more intuitive summary of how effective the optimization works.

\section{Experimental Results}
ChatSUMO with interactive web interface has been developed using Streamlit framework in Python. An example for simulation generation between user and Llama3.1 is visualized in Figure~\ref{fig:vehicle type change}. Furthermore, experiment study has been conducted leveraging the interface to evaluate the performance of ChatSUMO.

\subsection{Setup}
In the experiment section, we will conduct tests and experiments on the proposed ChatSUMO agent to demonstrate its effectiveness on simulation generation and modification. For testing based on LLM, we utilized Meta' s Llama3.1 for parsing text input by users. In the experimental part, we will evaluate its performance in two different construction of network, the abstract network and real-world network. As for the metrics, we focus on \textit{the average traffic density}, \textit{average travel time (TT)}, \textit{CO$_2$ emission} and \textit{fuel consumption (Fuel Cons)} as evaluation. As the distribution of vehicles in different level of roads varies a lot, to obtain the average traffic density, we summarize the top 10 congested roads' density and calculated the mean of summation.

\subsection{Simulation Generation}
\begin{figure*}[!htb]
    \centering
    \includegraphics[width=0.85\linewidth]{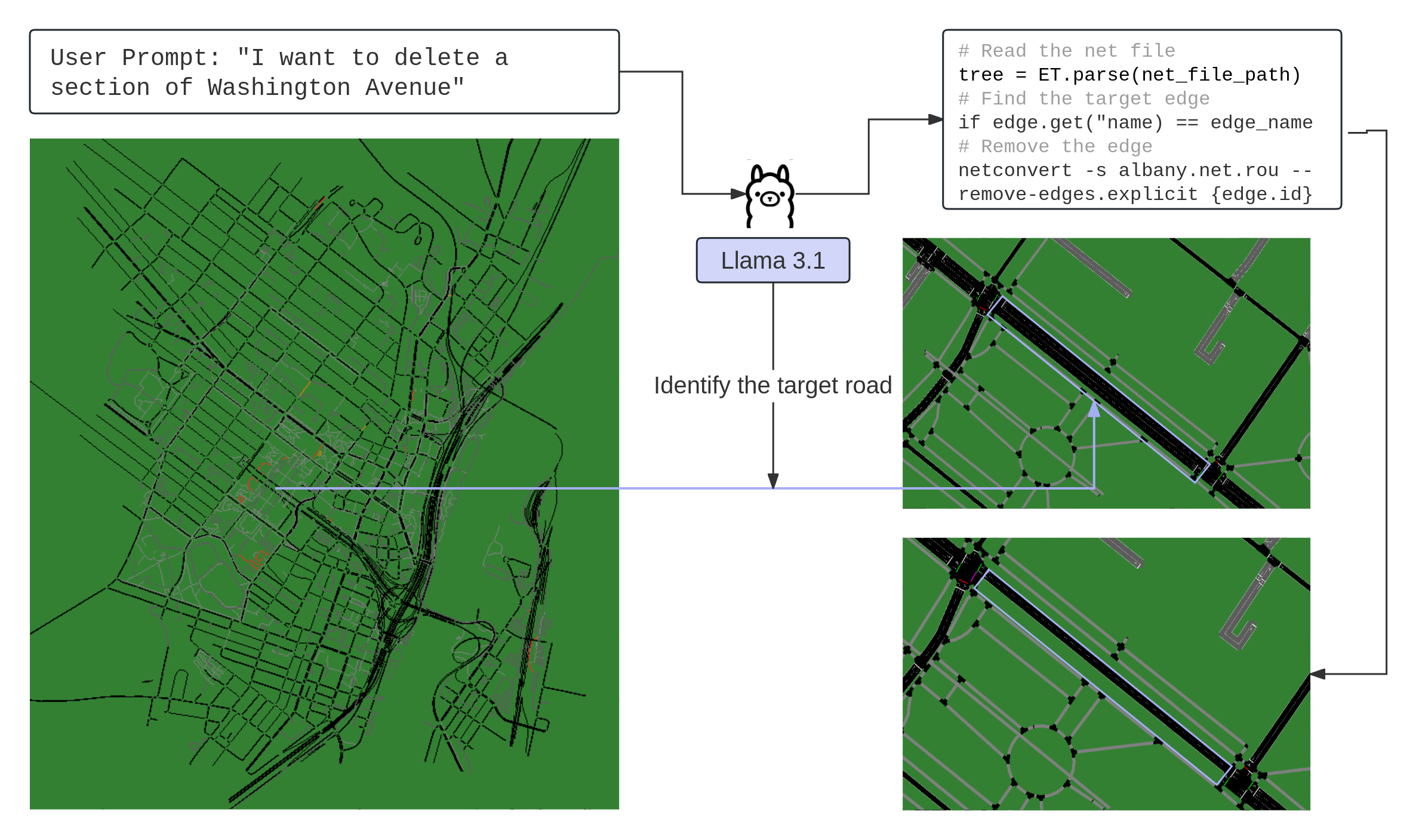}
    \caption{Edge Customization Experiment}
    \label{fig:edge experiment}
\end{figure*}

As the foundation of whole process, the accuracy of simulation generation plays an extremely important role in ChatSUMO. To evaluate the accuracy and effectiveness of simulation generation, we generate two types of networks. For the real-world network, we create a simulation of the city of Albany, New York with the radius of 1 miles around downtown. In order to make these two simulations relatively of the same size and same density of streets, we generate a spider-like network with 20 arms, 10 circles and the distance between circles is 150 meters. The setup for traffic condition for both simulations is ``medium", which is 2000 vehicles per hour. To meet these requirements, the user input is ``I want to see a traffic simulation in Albany. There should be medium traffic and it should show me streets in a 1 mile radius.". The generated real-world simulation is shown in Figure~\ref{fig:edge experiment}. To validate the accuracy of real-world network, we calculate the number of edges in the network created by ChatSUMO and the network downloaded by OSM. The number of the former is 30570 and the that of the latter is 29325, indicating that the difference is 4.2\%, which shows a great performance of generation module. To create such a simulation, it takes about one minute to complete the simulation, from entering user input, to generating the final summary. At the same time, we made a comparison with the time needed to build such a simulation manually which takes about 15 minutes. Considering the proficiency required for SUMO, beginners in traffic simulation may need to spend more time creating a complete traffic simulation, showing the important contribution of ChatSUMO in time efficiency.

To evaluate how well the system handles different scales of simulations, we conduct another experiment on recording the processing time for simulation generation in different scales, from small-scale intersections to city-wide traffic networks. In this experiment, we set the scale of network as three levels (0.5 mile, 1 mile, 3 miles) to simulate different traffic condition in city Albany. The experiment result is shown in Table~\ref{tab:simulation}. It can be observed that for small-scale intersections, the traffic simulation can be generated by ChatSUMO in 10 seconds, regardless the traffic condition. Regarding a normal scale, which is 1 mile, traffic simulation can be created in 30 seconds. However, the processing time of simulation increase significantly for a city-wide traffic simulation, considering the large scale network and large number of vehicles. In conclusion, the processing time for simulation generation depends on both the scale of simulation and the traffic condition, as well as the cpu performance of the conducted machine.

\begin{table}[!ht] \footnotesize
        \centering
        \renewcommand\arraystretch{1.5}
	\caption{Simulation Generation Experiment}\label{tab:simulation}
		\begin{tabular}{c|c c}
                \hline
			Traffic condition & Range (mile) & Processing Time (s) \\\hline
                \multirow{3}{*}{Medium} & 0.1 & 8.64 \\
                                      & 1 & 19.68 \\
			                       & 3  & 99.37 \\\hline
                \multirow{3}{*}{Heavy} & 0.1 & 9.49 \\
                                    & 1  & 23.38 \\
			                    & 3  & 174.3  \\\hline
		\end{tabular}
\end{table}

\subsection{Edge Modification}
Blocking some the of main streets has a significant impact in urban traffic, which would change the constriction of traffic flow. To assess the performance of edge editing, we implement the edit prompt in real-world network. In this experiment, we remove three levels of edges in simulation of Albany, which are Washington Avenue, Lark Street and Orange Street, to evaluate the impact of the edge removing function. At the same time, we utilize the customization in two different traffic conditions, with a volume of 2000 and 3000 vehicles per hour, to evaluate the modification impact differently. The visualization of modification is shown in Figure~\ref{fig:edge experiment}, which shows that our text commands successfully modified three type of streets in the simulated network.

The results of this experiment are shown in Table~\ref{tab:edge}. For the medium traffic, the removal of streets increases the average density of main streets, e.g. removing Washington Avenue lead to a increase of 3.32\% on density. At the same time, modification of streets changes the average travel time slightly. Removal of streets also boosts the CO$_2$ emission and fuel consumption. However, as the level of removed streets descends, the impact on different metrics gets smaller. For Heavy traffic, interestingly, the modification of streets, decreases the average density of main streets. The possible cause of this result is that in heavy traffic conditions, the density of main streets is already at a high level, and deleting an edge may lead traffic flow to another direction, decreasing the transit pressure for main streets. Unsurprisingly, the removal of roads also leads to a longer travel time in heavy traffic condition. Compared to average density and travel time, CO$_2$ emission changed significantly when vehicles increase from 2000 to 3000, with an increase about 53.1\%. In correlation with CO$_2$ emissions, more vehicles lead to obviously higher fuel consumption. In summary, removing lanes of different density levels affects traffic, but the lower the original density of the removed lanes, the smaller the impact on traffic.

\begin{table*}[!h] \footnotesize
        \centering
        \renewcommand\arraystretch{1.5}
	\caption{Edge Edit}\label{tab:edge}
		\begin{tabular}{c|c c c c c}
                \hline
			Traffic condition & Blocked Road & Density(veh/km) & TT(s) & CO$_2$ Emission(t) & Fuel Cons(t) \\\hline
                \multirow{4}{*}{Medium} & {Initial Network} & 195.90 & 287.62  & 1.60 & 0.51 \\
                                      & {Washington Avenue} & 202.48 & 291.44 & 1.63 & 0.52  \\
			                       & {Lark Street}       & 204.76 & 289.29 & 1.61 & 0.51  \\  
			                       & {Orange Street}     & 203.67 & 286.77 & 1.60 & 0.51  \\\hline
                \multirow{4}{*}{Heavy} & {Initial Network} & 220.61 & 293.06  & 2.44 & 0.78 \\
                                    & {Washington Avenue} & 212.06 & 302.69 & 2.52 & 0.80  \\
			                      & {Lark Street}       & 212.92 & 297.22 & 2.47 & 0.79  \\  
			                    & {Orange Street}     & 217.12 & 294.25 & 2.45 & 0.78  \\\hline
		\end{tabular}
\end{table*}

\subsection{Traffic Light Optimization}
\begin{figure*}
    \centering
    \includegraphics[width=1\linewidth]{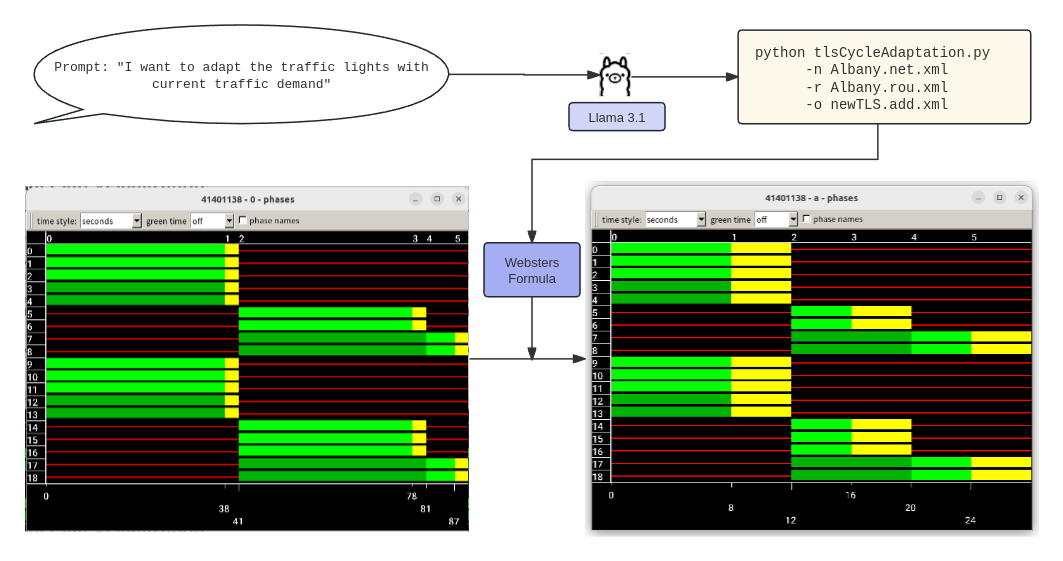}
    \caption{Traffic Light Adaptation}
    \label{fig:traffic light adaptation}
\end{figure*}

To optimize urban traffic signals, we have integrated two signal optimization modules in ChatSUMO: one for setting signal offsets and another for adjusting the duration of the green light phase, which are designed for both multiple and single traffic light optimization. Traffic light offsets is a powerful tools when dealing with multiple traffic light coordination in urban traffic, creating green wave for the crossing traffic flow and increasing the efficiency of transportation. To evaluate the impact of traffic light offsets on the two different network, in this experiment, we equip the simulations separately with traffic light offset with the command: ``I want to set traffic light offsets for the simulation''. To optimize individual traffic signals, the traffic light adaptation module in ChatSUMO can be utilized as shown in Figure~\ref{fig:traffic light adaptation}. The traffic signal program shown in the figure is implemented at the intersection of Madison Avenue and South Pearl Street. As we can see from the figure, after user input the prompt, the Llama3.1 process the text information and generate command creating the newTLS.add.xml file to modify the signal phases in SUMO simulation.

To verify the effectiveness of traffic light offsets and adaptation, we compare traffic flow density, travel times, CO$_2$ emission and fuel consumption (Fuel Cons) of the whole simulation. The experiment results are shown in Table~\ref{tab:traffic light}. We conducted tests and validations by utilizing the traffic light offset first and adapting traffic signals based on different traffic condition (medium and heavy traffic condition). In medium traffic condition, it is evident that traffic light offsets significantly decrease the average density of top 10 roads by 11.64\%, and they also reduce the average travel time by around 10 seconds. After utilizing traffic signal adaptation, however, the average density is even higher than the initial condition. On the contrast, the average travel time is reduced by surprisingly 40 seconds, which is 15.68\% shorter than the initial one. The probable explanation for the result is that the signal adaptation is designed to optimize a single intersection without considering the coordination of intersections. At the same time, CO$_2$ emission is decreased by 0.2t and fuel consumption is decreased by 0.06t.

\begin{table*}[!htb]\footnotesize
	\caption{Traffic Light optimization performance}\label{tab:traffic light}
        \renewcommand\arraystretch{1.5}
	\begin{center}
		\begin{tabular}{c|c c c c c}
            \hline
			 Traffic condition & Modification & Density(veh/km) & TT(s) & CO$_2$Emission(t) & Fuel Cons(t) \\\hline
                \multirow{3}{*}{Medium}  & Initial & 195.90 & 287.62  & 1.60 & 0.51 \\
                     & Traffic light offset        & 173.08 & 275.04  & 1.53 & 0.49 \\
                & Traffic light adaptation         & 199.13 & 242.53  & 1.39 & 0.44 \\\hline
			\multirow{2}{*}{Heavy}    & Initial & 220.61 &  293.06  & 2.44 &  0.78 \\
                            &Traffic light offset   &  205.37 & 315.23  & 2.60 &  0.83  \\
                        & Traffic light adaptation  & 225.75  & 246.20  & 2.12 &  0.68  \\\hline
		\end{tabular}
	\end{center}
\end{table*}

\subsection{Vehicle Edits}
\begin{figure*}
    \centering
    \includegraphics[width=0.85\linewidth]{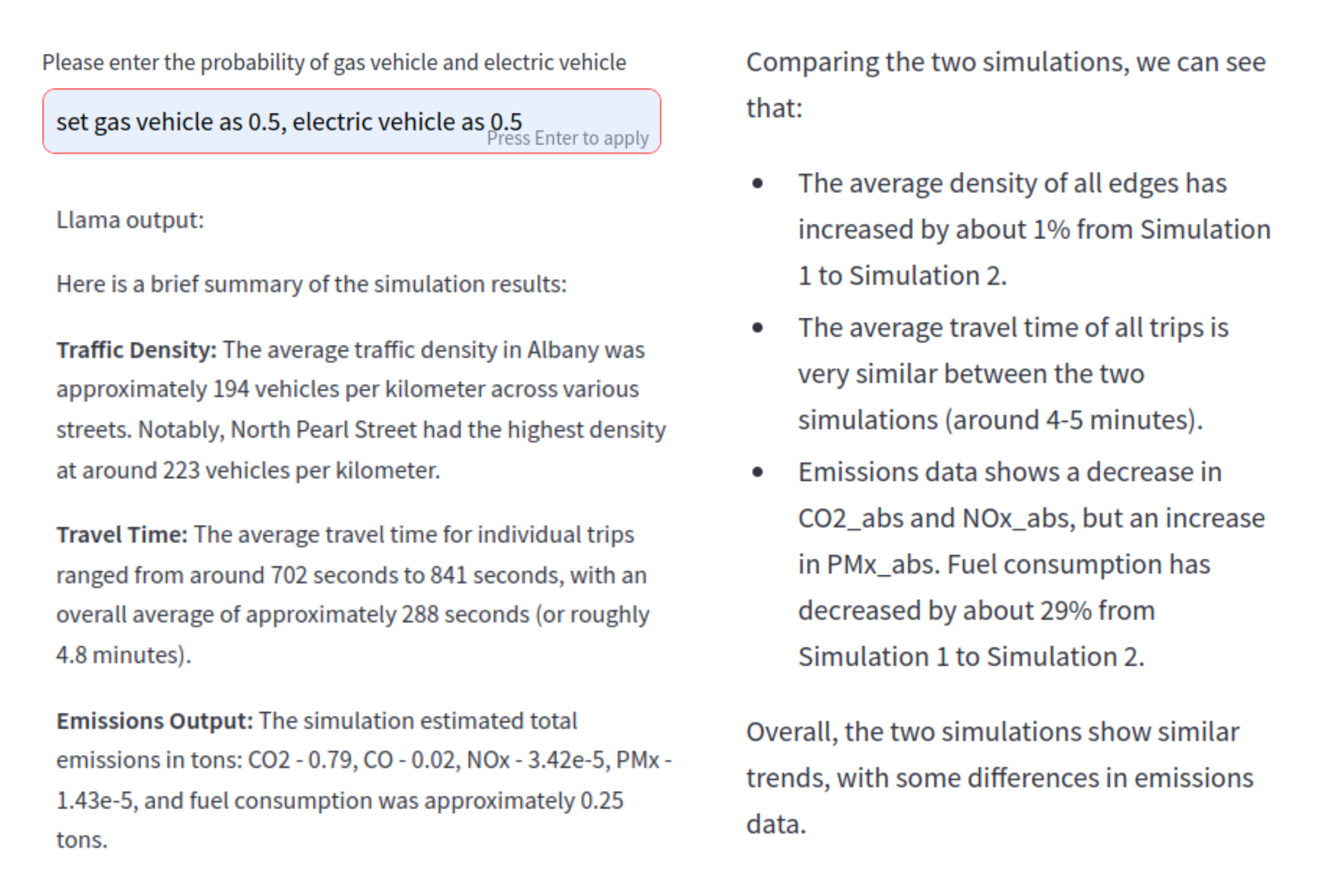}
    \caption{Vehicle Type Proportion Edit with the ChatSUMO Interface.}
    \label{fig:vehicle type change}
\end{figure*}

To compare the effectiveness of vehicle type customization, we prompt the proportion of electric vehicles from 0.3 to 0.5, aiming to observe the difference of pollutant emission and fuel consumption. Using the text input ``I want to set the proportion of electric vehicles as 0.5.'', we customize the vehicle type proportion. The output of both simulations are shown in Figure~\ref{fig:vehicle type change}. In these figures, we show the interactions with the ChatSUMO interface, and the output generated by Llama3.1 is also shown in three figures, which is very intuitive for users to see the summary of simulations. After implementing the vehicle type customization, the analysis module compares two simulations, and also generates a brief summary about general traffic, traffic density, pollutant emission and fuel consumption. It is obvious that the emission of $CO_2$ and fuel consumption has fallen by a very large amount with the increase of electric vehicles. However, the traffic density does not vary a lot due to the vehicle dynamic parameters are quite similar for both vehicle types.

\begin{figure*}
    \centering
    \includegraphics[width=0.85\linewidth]{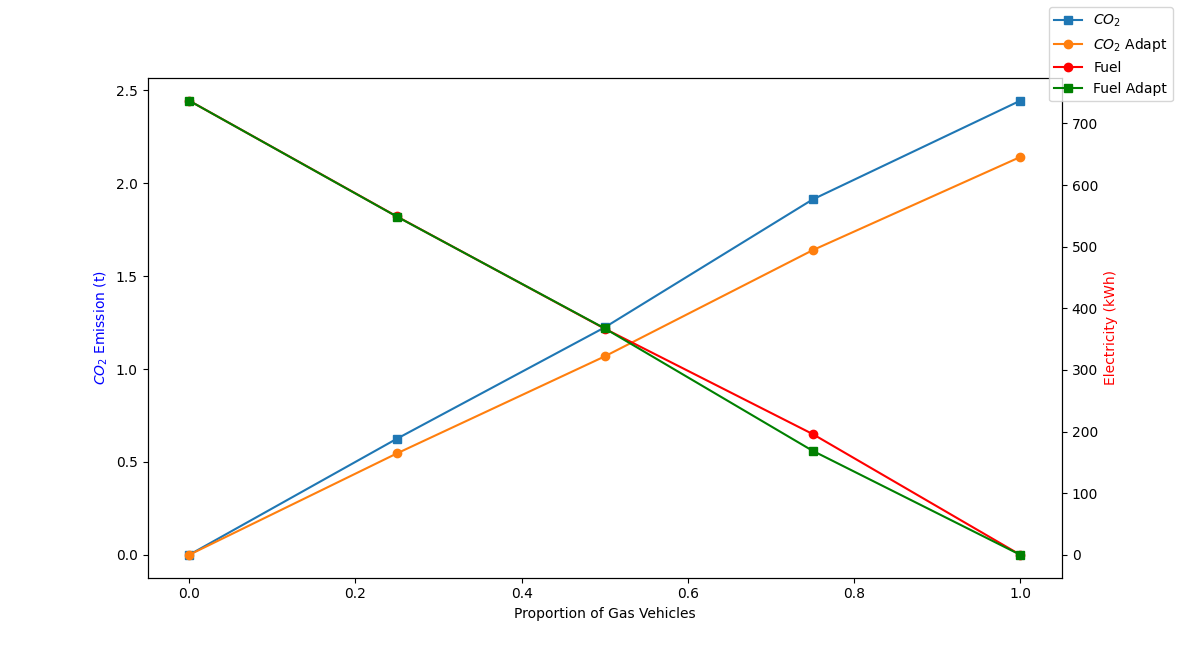}
    \caption{Emission trend with different vehicle proportion and traffic light adaptation supported by ChatSUMO}
    \label{fig:electricity change}
\end{figure*}

To evaluate the effectiveness of vehicle type editing and $CO_2$ and electricity trends with change of vehicle proportion by ChatSUMO and traffic light adaptation, we conducted an experiment with five different proportion of gasoline vehicle (0, 0.25, 0.5, 0.75, 1). All five simulation runs were automatically generated by ChatSUMO. The result of the experiment is shown in Figure~\ref{fig:electricity change}. It can be seen from the figure that $CO_2$ emission increases and electricity descents with the rises of gasoline vehicles' proportion. However, after utilizing the traffic light adaptation, although the emission of $CO_2$ decreases compared to the previous one, the electricity consumption is not affected according to the curve in the figure. We assume that the causing is the electricity consumption model is not sensitive to speed as travel distance.

\subsection{Discussion and Potential Application}
Through the experiments above, we have tested the ability of ChatSUMO in multiple fundamental functions, and the results of these experiments shows that ChatSUMO plays an active role not only in simulation generation but also in human and machine interaction. When conducting these experiences, thanks to ChatSUMO's excellent human-computer interaction experience, even though we made dozens of modifications to the simulation, the experiment itself did not take too long. Additionally, due to the involvement of the LLM, the results of each simulation were very intuitive, saving us a lot of unnecessary effort in our experiments.

The ease of use and excellent interactive experience of ChatSUMO provide it with great potential for application. For instance, ChatSUMO can be easily deployed as an online application, similar to ChatGPT, giving an approach for internet users to generate their own traffic simulation without mastering the conventional tools by SUMO. Users, especially beginning users can make preliminary testing on ChatSUMO by easily setting the simulation scenario by text massage, and then customize the scenario through some short words. Integration with real-world traffic, users can build the simulation with real-time traffic data through database API, and also simulate traffic incident like climate change or real-world road construction. With customized simulation and predefined metrics, users can do brainstorming for planning and estimating climate impacts as well. 

\section{Conclusion and Future Work}

In this paper, we have presented a comprehensive approach to generating SUMO simulations based on LLM. Our system, designed with the aim of democratizing access to traffic simulation tools, includes four key modules: user input, simulation generation, simulation modification, and output analysis. These modules work in concert to simplify the process of creating and refining traffic simulations, making it accessible to users with little to no prior experience in traffic modeling. The user input module ensures that users can easily specify their requirements and parameters without needing to understand the complexities of traffic simulation syntax. Through this integrated approach, we have demonstrated that complex traffic simulations can be generated, modified, and analyzed with minimal user intervention and expertise. The Llama3.1-based system not only reduces the barrier to entry for traffic simulation but also enhances the overall user experience by providing a seamless and intuitive interface. Future work will focus on further enhancing the system’s capabilities, including the incorporation of more advanced simulation features and improved user support tools, to continue expanding the accessibility and utility of traffic simulation technologies. To the best of our knowledge, we are the first to implement a large-language model with SUMO, integrating human understanding into simulation generation and modification. For the future work, we aim to generate more compre

\ifCLASSOPTIONcaptionsoff
  \newpage
\fi



%\bibliographystyle{IEEEtran}


% \end{thebibliography}
\bibliographystyle{IEEEtran}
\bibliography{refs}

\end{document}